\documentclass[11pt]{article}
\date{}

\usepackage{geometry}
\usepackage{amsmath,amsfonts,amssymb}  
\usepackage{mathtools}
\usepackage{bbold}

\usepackage{graphicx}		
\usepackage{subfigure}		
\usepackage{subcaption}		
\usepackage{float}			
\usepackage{placeins}

\usepackage{booktabs}		

\newtheorem{proposition}{Proposition}[section]

\usepackage[natbibapa]{apacite} 
\usepackage[hidelinks]{hyperref} 

\usepackage{cancel}		
\usepackage{comment}         
\usepackage[normalem]{ulem}  
\usepackage[dvipsnames]{xcolor} 

\usepackage{authblk} 

\title{Option-Implied Zero-Coupon Yields: Unifying Bond and Equity Markets}

\author[1*]{Ting-Jung Lee}
\author[1]{W. Brent Lindquist}
\author[1]{Svetlozar T. Rachev}
\author[2]{Abootaleb Shirvani}

\affil[1]{\small Department of Mathematics \& Statistics, Texas Tech University, Lubbock, TX 79409-1042, USA;
			brent.lindquist@ttu.edu, zari.rachev@ttu.edu}
\affil[2]{\small Department of Mathematical Science, Kean University, Union, NJ 07083, USA;
			ashirvan@kean.edu}
\affil[*]{\small Corresponding author, tinglee@ttu.edu}

\begin{document}
\maketitle
\begin{abstract}
This paper addresses a critical inconsistency in models of the term structure of interest rates (TSIR),
where zero--coupon bonds are priced under risk--neutral measures distinct from those used
in equity markets. 
 We propose a unified TSIR framework that treats zero--coupon bonds as European options with
deterministic payoffs ensuring that they are priced under the same risk--neutral measure
that governs equity derivatives.
Using put--call parity, we extract zero–coupon bond implied yield curves from S\&P 500 
index options and compare them with the US daily treasury par yield curves.
As the implied yield curves contain maturity time $T$ and strike price $K$ as independent  variables,
we investigate the $K-$dependence of the implied yield curve.
Our findings, that at--the--money, option--implied yield curves provide the closest
match to treasury par yield curves, support the view that the equity options market
contains information that is highly relevant for the TSIR.
By insisting that the risk--neutral measure used for bond valuation is the same as
that revealed by equity derivatives, we offer a new organizing principle for
future TSIR research.
\end{abstract}

\section{Introduction}\label{sec:Intro}
Modeling the term structure of interest rates (TSIR) is fundamental to financial theory and practice,
yet traditional TSIR models exhibit inconsistencies,
particularly in the valuation of zero--coupon bonds (zero bonds) under risk--neutral measures that differ
from measures used in equity markets.
This separation leads to distinct equivalent martingale measures (EMMs) for equities and fixed--income
securities,
despite the fact that both markets are inherently linked through investor risk preferences,
monetary policy, and market frictions.
As a result, existing models often fail to fully capture the interdependencies between bond and
equity pricing,
leading to systematic pricing discrepancies.
A substantial body of literature, focussed primarily on fixed--income markets, (Section~\ref{sec:lit rev})
has addressed the challenges of TSIR models from multiple perspectives.
Collectively these studies highlight persistent gaps between theoretical predictions and observed
yield dynamics.

The starting point for our analysis is founded on the first and second fundamental theorems
of asset pricing.
First, in a frictionless market with no arbitrage\footnote{
	More generally, in a market where there is ``no free lunch with vanishing risk''.
}
there exists at least one EMM under which the
discounted prices of all traded assets are martingales
\citep{Harrison_1979, Harrison_1981, Bjork_2009, Musiela_2005}.
Second, when the market is complete this measure is unique and is empirically identified by the
cross--section of traded contingent claims, in particular by liquid equity options.
Under this EMM, any European contingent claim is valued as the discounted conditional
expectation of its payoff.
A zero bond maturing at time $T$ with payoff \$1 is therefore just a special European claim,
and its price is obtained from the same risk-neutral measure that prices equity options.
This is exactly the logic in Shreve’s treatment of bonds and term--structure models
\citep[Chaps. 27, 28]{Chalasani_1997},
where the accumulation factor $\beta(t)$ is introduced and the bond price $B(t,T)$ is written
as the risk-neutral expectation of $1/\beta(T)$ conditional on the relevant filtration $\mathcal{F}(t)$.
Conceptually, the EMM is fixed once and for all by the global securities market;
it is not an object that can be independently chosen inside each asset class.

Under this perspective,  the TSIR must be derived using the same risk--neutral measure that
governs equity derivatives.
In macrofinancial language, one may think of the EMM as the ``risk--neutral world''
selected by a unique representative central bank that sets the short rate $r(t)$
as the risk--free return for all traded securities, regardless of their volatility.
Once this measure is determined by the prices of equity options in a complete market,
it must also govern the valuation of fixed--income securities, including zero bonds
of all maturities.
Allowing different EMMs for bonds and for equities would immediately open the door
to cross--market arbitrage, since one could replicate the same payoff using securities
priced under two distinct ``risk-neutral worlds''.

By contrast, the mainstream TSIR literature typically proceeds as if the fixed--income
market can choose its own EMM independently of the equity market.
Short--rate models such as those by \cite{Vasicek_1977, Cox_1985}; and \cite{Hull_1990},
as well as forward--rate and market models such as those by \cite{Heath_1992, Duffie_1996}; and
\cite{Brace_1997} begin by postulating dynamics for the short rate,
the entire forward--rate curve,
or a family of forward LIBOR rates under some reference measure.
An EMM is then constructed so that the discounted bond prices,
or the relevant forward rates, are martingales under this measure.
Within the fixed--income submarket, this construction yields an internally consistent,
arbitrage--free TSIR.
However, this procedure implicitly treats the bond market as if it were governed
by a central bank operating separately from  the equity options market.
The fundamental theorems of asset pricing are applied locally to the bond market,
whereas in an integrated capital market they should be applied globally to the
universe of equity, fixed--income, and derivative securities.

Our view is that this modeling practice creates a conceptual gap between
term--structure theory and the general arbitrage pricing paradigm.
Once the equity derivatives market is complete and liquid,
the EMM is no longer a free modeling input for TSIR models;
it has already been identified by the prices of European options on the underlying
equity index.
In particular, the zero bond that pays \$1 at time $T$ should be priced under
exactly the same EMM as any other European claim, and its yield should therefore
be consistent with the option-implied distribution of the underlying equity index
over $[0,T]$.
In this sense, we argue that much of the TSIR literature has, perhaps inadvertently,
``forgotten'' the global nature of the risk--neutral measure emphasized in the foundational
work on arbitrage pricing \citep{Harrison_1981, Shreve_2004II}.

A parallel inconsistency appears in discrete--time models.
In binomial term--structure models and related interest--rate trees,
it is common to choose risk--neutral up and down probabilities in an ad hoc manner,
often setting them equal to one half for convenience,
and then to build bond and interest--rate dynamics under this artificially chosen
risk--neutral measure \citep[see, for example][]{Shreve_2004I}.
\cite{Hu_2020} have shown that such symmetric risk--neutral probabilities are not supported by the
empirical distribution of underlying returns and that they are statistically
misspecified when confronted with equity and option data.
The deeper message is the same: once the EMM has been fixed by traded option prices
in a complete market, it should be used consistently across all derivative and
fixed--income contracts rather than being redefined within each separate modeling context.

The present paper is intended to be “ground breaking” in precisely this conceptual sense.
We close the risk--neutral measure gap between the equity and bond markets by treating
zero bonds explicitly as European options with deterministic payoffs and by
deriving their prices directly from equity option data.
Using put--call parity (PCP) on S\&P 500 index options, we construct implied yield curves
for zero bonds and compare these curves with US daily treasury par yield curves.
Because our construction is anchored in the equity options market,
the resulting TSIR is, by design, evaluated under the same EMM that governs equity
derivatives.
This unifies the valuation of fixed--income and equity derivative securities within
a single arbitrage--free framework and eliminates the artificial separation created
when TSIR models choose their own bond--market EMM.
Conceptually, our approach restores the spirit of the fundamental theorem of asset
pricing to term--structure modeling: there is one risk--neutral world,
calibrated in practice from equity options,
and all zero bonds — hence the entire term structure — must be priced in
that world.

This paper is structured as follows.
Section~\ref{sec:lit rev} reviews the literature on the TSIR and PCP violations.
Section~\ref{sec:two_forms} presents the two competing
no--arbitrage formulations of TSIR and discusses their theoretical consistency.
Section~\ref{sec:empirical} presents our empirical analysis.
As noted above, we define the implied yield of a zero bond using the European option PCP equation.
We utilized cash-settled call and put options on the S\&P 500 index \^{}SPX,
as this provides an underlying which represents a broad equity market.
We compared the implied yield to daily treasury  par yield curves.
Unlike the par yield curves, which are functions of maturity time $T$ only,
the option--implied yields depend on $T$ and strike price $K$.
One can think of the implied yields as a family of curves $Y^{(\text{imp})}(t,T;K)$
parameterized by a strike value (equivalently by a moneyness value $M = K/S_t$,
where $S_t$ is the value of \^{}SPX at time $t$).
Using box-whisker summaries of distributions, we examined the variation of this
family of implied curves, finding that in--the--money and out--of--the--money values of
$K$ tended to increase deviations of the implied yield from the corresponding treasury par yield,
whereas at--the--money options provided much closer agreement.
Section~\ref{sec:concl} concludes with a discussion of our results.

By no means do we claim that the empirical approach adopted here is definitive.
Rather, at the empirical level the unified EMM perspective enables us to test whether
option--implied, zero bond yields are consistent with observed treasury yields
and to quantify the extent of any residual discrepancies.
Our findings, that at--the--money option--implied yield curves provide the closest
match to treasury par yield curves, support the view that the equity options market
contains information that is highly relevant for the TSIR.
More broadly, by insisting that the EMM used for bond valuation is the same as
that revealed by equity derivatives, we offer a new organizing principle for
future TSIR research.
Existing short--rate, forward--rate, and market models can be reexamined and
recalibrated under the equity--implied EMM, rather than under a bond--specific measure,
thereby reestablishing the coherence of asset pricing across markets.

\section{Literature Review} \label{sec:lit rev}
The modeling of the TSIR has been one of the central topics in financial economics
since the seminal contributions of \cite{Vasicek_1977} and \cite{Cox_1985}.
These early short--rate models provided tractable frameworks for bond pricing but
struggled to match the observed dynamics of the yield curve.
A significant, arbitrage--free, TSIR model framework was introduced by  \cite{Heath_1992}.
In this approach, modeling begins with the dynamics of forward rates and derives the
EMM under which the entire family of zero bonds is arbitrage free.
\cite{Duffie_1996, Miltersen_1997, Brace_1997} derived models with further enhancements under
this TSIR framework.
\cite{Fisher_2004}  presented a foundational overview of discrete TSIR modeling approaches.

While these models secured internal consistency, they often relied on assumptions that created
discrepancies when compared with equity--based, risk--neutral valuation.
The equity--based measure reflects probabilities embedded in option prices,
whereas the bond market embeds its own risk--neutral measure inferred from treasury yields
and interest--rate derivatives.
From our perspective, \cite{Duffie_2001} and \cite{Shreve_2004II} significantly advanced the foundations of
asset pricing by establishing that zero bonds should be valued under the same risk--neutral measure
governing equity options.

Despite this, most TSIR models continued to employ distinct EMMs for fixed--income and equity markets,
resulting in inconsistencies across asset classes.
The focus on a TSIR framework divorced from equity markets has resulted in the developments of hypotheses
related to bond prices and the TSIR yield curve.\footnote{
	For US--based studies, the yield curve is usually interpreted to be that of US government treasury
	securities.
}
The expectations hypothesis \citep{Gurkaynak_2012} states that long--term yields are equal to the average
of expected short--term interest rates until the (long--term) maturity date.
(That is, long--term bond rates are solely determined from current and future short--term rates.)
In a more expansive form, this has also been referred to as the spanning hypothesis, which
proposes that the yield curve spans all information relevant for forecasting future yields and returns;
no variables other than those embodied in the current yield curve are needed for forecasting \citep{Bekaert_2021}.
This hypothesis, in either form, has proved troubling as empirical evidence \citep{Cochrane_2005, Sarno_2007}
demonstrated that  that expectations--based models systematically failed to capture the dynamics of both
short-- and long--term yields.
This work pointed to the importance of  time--varying risk premia and suggested that the TSIR is likely
to be ``considerably more complex'' than that suggested by the expectations hypothesis.

A number of studies have established empirical links between macroeconomic factors and the TSIR
\citep{Ang_2003, Ludvigson_2009, Cooper_2009, Bikbov_2010, Joslin_2014, Cieslak_2015}.
This has coincided with the development of macrofinance term structure models (MTSMs) in either, so--called,
reduced or equilibrium forms.
(See \citet[Footnotes 1 and 2]{Bauer_2017} for references.)
In contrast, \cite{Duffee_2013} argued that ``none of our models [linking TSIR with the macroeconomy]
is consistent with basic properties of nominal yields'',
while \cite{Bauer_2017} and \cite{Bauer_2018} cited serious small--sample distortions in such studies
and suggested the evidence against the spanning hypothesis ``is much weaker than it originally appeared''.
Subsequent model development by \cite{Bekaert_2021} ``resurrected the evidence against the spanning
hypothesis''.

Collectively, this stream of literature highlights both theoretical and empirical tensions in the valuation
of fixed--income securities.
While not questioning the importance of this work,
we note that such TSIR models emphasize no--arbitrage conditions under bond--specific measures,
while asset pricing theory indicates that a unified risk-neutral measure should prevail across equity and
fixed--income markets.

Implicit in our empirical work in Section~\ref{sec:empirical} is that any PCP violations should be
infrequent and small in magnitude.
Summarizing existing literature results, \cite{Kamara_1995} noted that empirical studies reported
``frequent, substantial violations'' of PCP in American options.\footnote{
	For American options,  PCP consists of lower and upper bound inequalities
	but can be rewritten as an equality by introducing early exercise premium terms for
	the call and put options).
}
\cite{Ofek_2004} examined the effect of short sales restrictions on PCP violation
of American options and
\cite{Nishiotis_2019} highlighted how regulatory constraints and short--sale restrictions amplified
violations for American options during the 2008 the US short--sale ban.

In contrast to American options, \cite{Kamara_1995} showed that PCP violations
are ``less frequent and smaller'' for (cash-settled) European options  on the S\&P 500 index and
are primarily due to ``the liquidity risk of adverse price movements from order submission until
order execution''.
An empirical study by \cite{Chesney_1995} of index options on the Swiss Market Index
found severe PCP violations in that relatively young option market, with the violations decreasing
over the study time period, which the authors attributed to a learning curve by market participants.
They cited transaction costs, short selling restrictions and the impact of block trades as partial
explanations for the persistence of violations.
\cite{Mittnik_2000} simulated trading strategies for exploiting PCP violations for
(the then relatively new) European options on the German DAX index.
The simulations incorporated transaction costs and execution delays, two market frictions that
limit arbitrage.
In general they found that violations occur with statistical significance, but are not economically
exploitable.
As these index options were relatively new, the authors were able to show that, as traders and
market makers became more experienced with the new instrument, violations became less
frequent.
More recently, \cite{Steurer_2022} examined PCP violations for European options
on the Shanghai Stock Exchange (SSE) 50 ETF which tracks the SSE 50 index.
With the qualification that their study covers only a short time period,
the authors concluded that arbitrage opportunities are ``literally not existent''.

We consequently proceed with the assumption that PCP violations of cash-settled,
European index options in the efficient US market are infrequent and small in magnitude.

\section{Two Formulations of Arbitrage--Free TSIR in a Complete Market Model}\label{sec:two_forms}
We illustrate the discrepancy in risk--neutral formulations through two propositions
both of which assume a fair (arbitrage--free) valuation of the TSIR.

Consider the stochastic basis
$(\Omega, \mathbb{F} = \{ \mathcal{F}_t,t \ge 0\} \subset \mathcal{F}, \mathbb{P})$
on a complete probability space $(\Omega, \mathcal{F}, \mathbb{P})$.
Let $B_t$, $t \ge 0$ denote a standard Brownian motion defined on $(\Omega, \mathcal{F}, \mathbb{P})$
with the filtration $\mathcal{F}_t = \sigma(B_u, 0 \le u \le t)$, $t \ge 0$. 
We consider a market $(\mathcal{S}, \mathcal{B},\mathcal{C})$ consisting, respectively,
of a risky asset, a riskless asset and a European contingent claim (option).
We assume the price  $S_t$ of the risky asset
follows a continuous stochastic process with dynamics given by\footnote{
	The regularity conditions for ${\mu}_t, t \ge 0$ and ${\sigma}_t, t \ge 0$ are described in
	\citet[{Appendix E}]{Duffie_2001}.
}
\begin{equation} \label{eq:dSt}
	dS_t = \mu_t S_t dt + \sigma_t S_t dB_t, \qquad t \ge 0, \ \mu_t \in R, \ \sigma_t > 0, \ S_0 > 0.
\end{equation}
The price $\beta_t$ of the riskless asset  evolves as
\begin{equation} \label{eq:dBt}
	d\beta_t = r_t \beta_t dt, \qquad \beta_0 > 0.
\end{equation}
{With the appropriate regularity conditions\footnote{
	The regularity conditions for $r_t, t \ge 0$ are described in \citet[Section 5G]{Duffie_2001}.
}
}
on $r_t$, \eqref{eq:dBt} has the solution
\begin{equation}\label{eq:Bt}
	\beta_t = \exp \left\{ \int_0^t r_s ds \right\}.
\end{equation}
We assume $\mu_t$, $\sigma_t$ and $r_t$ are $\mathcal{F}$--adapted.
The price $C_t$ of the option has the dependence $C_t = f(S_t,t)$, $t \in [0,\tau]$,
with terminal payoff at the maturity time $\tau$, given by $C_\tau=g(S_\tau)$.\footnote{
	The regularity conditions for $f(x,t)$, $x>0$, $ t \in [0, \tau)$, and $g(x)$, $x>0 $
	are described in \citet[Section 5G]{Duffie_2001}.
}

The market $(\mathcal{S}, \mathcal{B},\mathcal{C})$ is arbitrage--free and complete provided
the market price of risk
$\theta_t = (\mu_t - r_t)/\sigma_t$ satisfies the condition
\begin{equation}\label{eq:mpor}
    \theta_t = \frac{\mu_t - r_t}{\sigma_t} > 0, \quad \mathbb{P} \text{-a.s. for all } t \geq 0.
\end{equation}
As a consequence, there exists \citep[][Sections 6G and 6H]{Duffie_2001} a unique EMM 
$\mathbb{Q}$ such that the standard Brownian motion $B_{t}^{\mathbb{Q}}$ on
$(\Omega,\mathbb{F},\mathbb{Q})$  is related to the natural--world Brownian motion $B_t$
through the market price of risk,
\begin{equation}\label{eq:BtQtoP}
	dB_{t}^{\mathbb{Q}} = dB_t + \theta_t dt.
\end{equation}
The (risk--neutral) dynamics of $\mathcal{S}$ on $(\Omega,\mathbb{F},\mathbb{Q})$ is given by
\begin{equation}\label{eq:rndSt}
	dS_t = r_t S_t dt + \sigma_t S_t dB_{t}^{\mathbb{Q}}, \quad t \in [0,\tau].
\end{equation}
The risk--neutral (fair) value of the contingent claim $C$ at any time $t \in [0,\tau)$ is given by the
standard risk--neutral valuation formula \citep[][p116, equation (15)]{Duffie_2001}
\begin{equation}\label{eq:rnCt}
	C_t = \mathbb{E}_t^{\mathbb{Q}} \left[ e^{-\int_t^\tau r_s ds} g(S_\tau) \right].
\end{equation}
The first proposition forms the basis for risk--neutral asset valuation of the TSIR in the
market model defined by \eqref{eq:dSt}--\eqref{eq:rnCt}.

\begin{proposition}\label{prop:BSM_TSIR}
Consider a unit zero bond $\mathcal{Z}_t$, for $0 \leq T \leq \tau$,
which can be viewed as an option with maturity $T$ and payoff $g(\mathcal{Z}_T) = \$1$.
From \eqref{eq:rnCt}, the risk--neutral value of $\mathcal{Z}_t$ for $t \in [0,T]$ is given by 
(see also, \citet[p127, equation (21)]{Duffie_2001} or \citet[p267]{Chalasani_1997})
\begin{equation}\label{eq:LamtT}
	\Lambda(t,T) = \mathbb{E}_t^{\mathbb{Q}} \left[ e^{-\int_t^T r_s ds} \right],
	\qquad t \in [0,T].
\end{equation}
{\rm The collection $\Lambda(t,T)$ or, equivalently, $Y(t,T) = \ln \Lambda(t,T)$, for $0 \leq t \leq T \leq \tau$,
defines the TSIR.}
\end{proposition}

The alternate proposition considers the $\mathcal{F}$--adapted interest rate process $r_t$ defining
the riskless asset price \eqref{eq:Bt} for $t \in [0,T]$.
It assumes the family of all unit zero bonds $\mathcal{Z}_T$ having maturities over the range
$0 \leq T \leq \tau$ are default--free.
Let
$$
	W(t,T) \text{ denote the price of the zero bond } \mathcal{Z}_T \text{ at time } t \in [0,T].
$$
It is assumed that $W(t,T)$  is adapted to the filtration $\mathbb{F}_{[0,T]} = \{ \mathcal{F}_t,t \in [0,T]\}$.
The alternative TSIR (ATSIR) is defined as the collection $\{W(t,T),\ 0 \leq t \leq T \leq \tau\}$.

\begin{proposition}\label{prop:Alt_TSIR}
The ATSIR is free of arbitrages if and only if there exists a probability measure\footnote{
	$\tilde{\mathbb{P}} \sim \mathbb{P}$ states $\tilde{\mathbb{P}}$ and $\mathbb{P}$
	are equivalent probability measures; specifically, for $A \in \mathcal{F}$,
	$\tilde{\mathbb{P}}(A) = 0$ if only if $P(A)  = 0$.
}
$\tilde{\mathbb{P}} \sim \mathbb{P}$ such that  the discounted zero bond price
\begin{equation}\label{eq:WtT}
	\frac{W(t,T)}{{\beta}_t}, \qquad t \in [0,T]  
\end{equation}
is a martingale under the probability space $(\Omega, \mathbb{F}_{[0,T]},\tilde{\mathbb{P}})$.
\end{proposition}
As noted in the introduction, the alternate proposition has been widely applied in TSIR
modeling (see e.g.,
\citet[Chapter 10]{Shreve_2004II}, \citet[Chapters 4--6]{Cairns_2004} and \citet[Chapter 4]{Wu_2019}).

As the market $(\mathcal{S}, \mathcal{B},\mathcal{C})$ is assumed to be complete and arbitrage free,
and $\mathcal{C}$ can represent any zero bond $\mathcal{Z}_T$,
from the point of view of Proposition~\ref{prop:BSM_TSIR}
it follows that $\tilde{\mathbb{P}}$ must coincide with $\mathbb{Q}$.
However, Proposition~\ref{prop:Alt_TSIR} claims that $\tilde{\mathbb{P}}$ is fundamentally different
from $\mathbb{Q}$, leading to inconsistent views.
Proposition~\ref{prop:Alt_TSIR} basically states that there exist two central banks,
one dealing with equity pricing and all options, and another just for the zero bonds.
The measure $\mathbb{Q}$ is derived from the market price of risk \eqref{eq:mpor} and is influenced
by the volatility $\sigma_t$, which is a key component of asset pricing models.
In contrast, the measure $\tilde{\mathbb{P}}$ is defined independently of $\theta_t$ and,
more critically, is indifferent to the volatility $\sigma_t$ of the equity market.\footnote{
	Equity market volatility can indeed impact bond yields, as demonstrated by various sources.
	\begin{enumerate}
	\item The Federal Reserve Bank of St. Louis highlighted that equity market volatility,
        		tracked by the VIX and realized volatility of returns on the S\&P 500,
        		moves in tandem with bond yields.
        		Increased equity volatility often leads investors to shift their investments to bonds,
        		impacting the yield curve and bond performance
        		(\url{https://fred.stlouisfed.org/series/EMVMACROINTEREST}).
	\item The CME Group discussed how equity market conditions and the actions of the
		Federal Reserve, such as quantitative easing, influence bond yields.
		They noted that periods of high equity volatility typically result in lower bond yields
		due to increased demand for safer assets
		(\url{https://www.cmegroup.com/education/featured-reports/end-of-the-road-for-the-yield-curve-vs-volatility-cycle.html}).
	\item The Office of Financial Research elaborated on how rising interest rates and
		equity market volatility pose risks to insurers' investment portfolios.
		They indicated that insurers exposed to equity market volatility might face reduced liquidity
		and increased risk, affecting their bond investment strategies and yields
		(\url{https://www.financialresearch.gov/the-ofr-blog/2022/07/21/Rising-Interest-Rates-Help-Insurers-but-Market-Volatility-Poses-Risk-to-Some/})
	\end{enumerate}
}

\section{Empirical Comparison of Yield}\label{sec:empirical}
The conclusion that Proposition~\ref{prop:BSM_TSIR} provides the appropriate universal
valuation of a TSIR, provides the motivation to compare the yield determined by the
proposition with observed market yields.
We define an implied yield based on the
parity relationship for European put--call pairs of option contracts,
specifically of  cash-settled options on the broad market index \^{}SPX.
The specific option chains used were for the five trading dates in the period
09--15 October, 2024.\footnote{
	Data source: Yahoo Finance. Accessed 16 October, 2024.
}
The option chains were based upon close--of--market values of the index.
We took the market yield to be the rates of the US daily treasury par yield curve
for the same five trading days.\footnote{
	Data source: US Treasury, accessed June~6,~2025.
	As 14 October 2024 was a federal holiday, and as the par yield rates had small daily changes,
	we obtained equivalent market rates for 14 October by simply averaging those for 11 and 15 October. 
}
We examined the dependence of the implied yield on maturity time and strike price.
We defined two measures of dislocation between the implied yield and
the par yield curve and examined this dislocation as a function of the time to  maturity.

\subsection{Implied Yield as a Function of Strike and Maturity}\label{sec:ImpY} 
We define the discount factor  of a zero bond (Proposition~\ref{prop:BSM_TSIR}) using
the European option PCP equation
\begin{equation}\label{eq:pcp}
	\Lambda(t, T, K) = \frac{1}{K} \left[ S_t + P(t, T, K) - C(t, T, K) \right],
\end{equation}
where
$P(t, T, K)$ and $C(t, T, K)$ are, respectively, put and call option prices for contracts
having maturity $T$ and strike price $K$,
and $S_t$ is the spot price of the underlying asset
(in our case the value of the index \^{}SPX).
As a consequence of this definition, the discount factor $\Lambda(t, T, K)$ is dependent on
$T$ and $K$.
The $T, K$--dependent yield implied from \eqref{eq:pcp} is
\begin{equation}\label{eq:pcpY}
	Y(t, T, K) =  - \frac{1}{T - t} \ln[\Lambda(t, T, K)].
\end{equation}
We refer to $Y(t, T, K)$ as the ``option implied yield'' (implied yield for brevity).
We use the notation $Y^{\text{mkt}}(t,T)$ to refer to the rates from the treasury par
yield curves.
From \eqref{eq:pcpY} and \eqref{eq:pcp}, and noting that $S_t$, $K$, $P(t, T, K)$ and $C(t, T, K)$
are always positive, we have\footnote{
	Technically, \eqref{eq:negYa} should be written $ C(t, T, K) < S_t + P(t, T, K) < K + C(t, T, K)$,
	but as $C(t, T, K) < S_t$ always (otherwise why purchase the call option?),
	we have neglected the first inequality.
}
\begin{subequations}
\begin{align}
	Y(t, T, K) > 0 \text{ if }  0 &< \Lambda(t,T,K) < 1 \nonumber\\
		&\rightarrow  S_t + P(t, T, K) < K + C(t, T, K) , \label{eq:negYa}\\
	Y(t, T, K) < 0 \text{ if }  1 &< \Lambda(t,T,K) \nonumber\\
		&\rightarrow S_t +  P(t, T, K)  > K + C(t, T, K). \label{eq:negYb}
\end{align}
\end{subequations}

\begin{figure}[h!]
	\centering
	\includegraphics[width=0.49\textwidth]{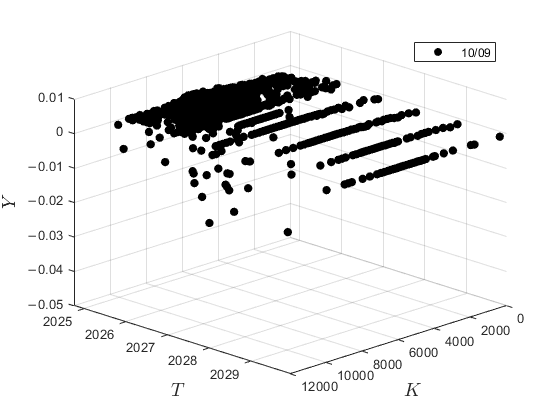}
	\caption{3D scatter plot of implied yields $Y(t,T,K)$
		 versus strike $K$ and maturity $T$ for the trading day $t =$ 09 October, 2024.
	}
	\label{fig:Y3D}
\end{figure}

Fig.~\ref{fig:Y3D} presents the implied yield $Y(t,T,K)$ computed from the option chain of 09 October, 2024.
Fig.~\ref{fig:Y3D_rest} in the supplementary material presents the implied yields computed for the option chains
of 10--15 October, 2024.
The plot highlights two key features.
First, the implied yield surface appears to be centered close to zero,\footnote{
	However, the $y-$axis scale of $O\left(10^{-2}\right)$ is quite large.}
consistent with PCP under no--arbitrage.
Second, large, negative values of $Y(t,T,K)$ occurred more frequently for $K < S_t$
(i.e., for $M = K / S_t < 1$) 
and short horizons (e.g., $T-t \lesssim 45$ days).
Writing \eqref{eq:negYb} as $(S_t-K) +  P(t, T, K)  > C(t, T, K)$, 
we suspect \eqref{eq:negYb} is satisfied for $K < S_t$ when  a current value of $S_t-K$ is compared to
either a stale put or call price on a $T,K$ contract at short maturity times.
Yield values stabilize closer to zero as the time to maturity increases.

\begin{figure}[h!]
	\centering
	\includegraphics[width=0.5\textwidth]{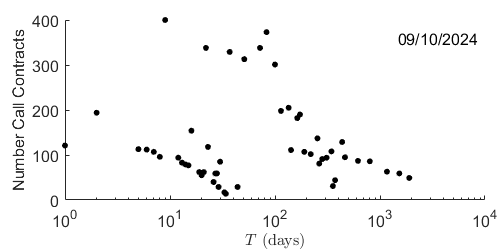}
	\caption{Two-dimensional scatter plot of the number of \^{}SPX call contracts versus time
		to maturity $T$ for the option chain of 09 October, 2024.}
	\label{fig:contracts_1009}
\end{figure}
Our data set consists of a mixture of contract types: daily, monthly, quarterly and longer--term.
Fig.~\ref{fig:contracts_1009} displays a two dimensional scatter plot of the number of \^{}SPX
call contracts versus time to maturity $T$ for the option chain of 09 October, 2024. 
The distribution of contracts exhibits two distinct clusters:
a cluster of contracts consisting of short maturity ``dailies'' (extending out to $T = 44$ days)
and a cluster extending over longer horizons, which includes monthly, quarterly and
long--term equity anticipation securities.
We refer to the two clusters as ``short term'' and ``long term''.
This clustering pattern was also observed for the put options and appeared in each of the
four option chains studied over the period 10--15 October, 2024
(see Fig.~\ref{fig:contracts_rest} in the supplementary material).
To look for potential differences in the implied yield $Y(t,T,K)$,
we analyzed short-- and long--term contracts separately.

\begin{figure}[h!]
	\centering
	\includegraphics[width=\textwidth]{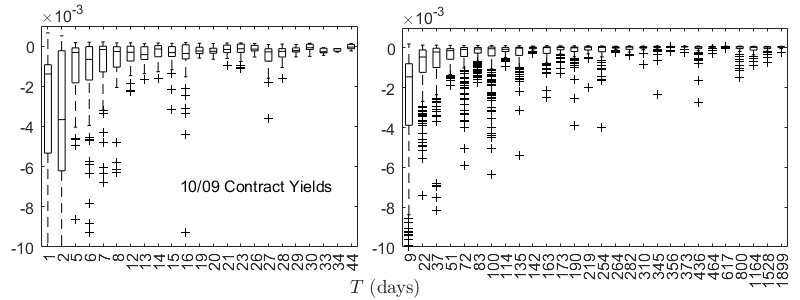}
	\caption{Box-whisker summaries of the distributions of the implied yield $Y(t,T)$ with maturity
		time $T$. The panels summarize (left) short--term and (right) long--term contracts.
		For better representation, the negative range of the $y-$axis has been truncated;
		consequently, negative outliers for small values of $T$ are not visible.
	}
	\label{fig:YTB}
\end{figure}

Fig.~\ref{fig:YTB} presents box-whisker summaries of the variation of the distribution of $Y(t,T,K)$
with strike price $K$ at each fixed maturity time $T$ for the data in Fig.~\ref{fig:Y3D}.
The short-- and long--term contracts are summarized separately.
(Fig.~\ref{fig:YTB_rest} in the supplementary material  provides the same summaries for the option chains
of 10--15 October, 2024.)
For fixed $T$, we note that  the distribution is generally skewed towards negative yield values,
with the skewness most prevalent at the shorter maturity times.
The number of negative yield outliers is more predominant in the long--term contracts.
There is  sizable variation in yield values at smaller maturity values, with the variation
significantly diminishing as maturity time increases.
The predominance of negative implied yields computed using PCP cannot be ascribed simply
to effects such as stale call or put prices due to asynchronous trading of contracts.
However, our interest is not in the reason for the negative yields, but rather on effective
measures of the discrepancy between implied and published yield rates.
To compare the implied yield rates $Y(t,T,K)$ against a published rate $Y^{(\text{mkt})}(t,T)$ for
day $t$ requires a method of aggregating the implied rates.
In sections~\ref{sec:medY} and \ref{sec:ATM}, we consider two aggregation methods.

\subsection{Median-Strike Aggregation of Implied Option Yields} \label{sec:medY}

We considered a ``median strike'' method,
where the dependence of the implied yield on the strike price is reduced to a median value
at each fixed value of $T$.
Specifically, from \eqref{eq:pcpY} we computed\footnote{
	The median was chosen instead of the mean as it is unaffected by
	the magnitude of outlier values.
}
\begin{equation}\label{eq:impY}
	Y^{(\text{med})}(t,T) =  \underset{K}{\text{median}}\{ Y(t, T, K) \}.
\end{equation}
We define the maturity--dependent dislocation between the yield \eqref{eq:impY}
and the market yield by
\begin{equation}\label{eq:DY}
	\Delta_Y^{(\text{med})}(t,T) =  Y^{(\text{med})}(t,T) - Y^{(\text{mkt})}(t,T).
\end{equation}

\begin{figure}[h!]
	\centering
	\includegraphics[width=0.35\textwidth]{yieldcurve_20241009.png}
	\caption{PCHIP fits to the 09 October 2024, US daily treasury par yield curve rates for maturities
		$T$ of 1 month through 30 years.
		The fit for $T < 1$ month uses the rate for $T = 1$ month.
	}
	\label{fig:pchip}
\end{figure}

The treasury par yield curve bill, note and bond rates have maturities $T \in \{1,2,3,4,6\}$ months
and $T \in \{ 1,2,3,5,7, 10, 20, 30\}$ years.
As the maturity time of the option chains do not usually align with the par yield curve maturities,
we interpolated the daily par yield curve rates using a piecewise cubic Hermite interpolating polynomial (PCHIP)
to provide a market rate for any option maturity time between one month and 30 years.
For option maturity times below one month, we simply used the one month par yield rate.
Fig.~\ref{fig:pchip} shows the PCHIP fit to the par yield curve rates for 09 October 2024.
The fits to the par yield curve rates for the other four trading days are very similar to that shown in
Fig.~\ref{fig:pchip}.

\begin{figure}[htb]
	\centering
	\includegraphics[width=0.3\textwidth]{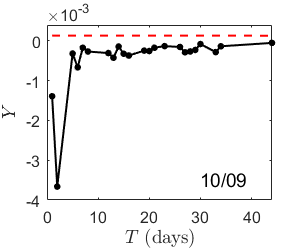}
	 \includegraphics[width=0.3\textwidth]{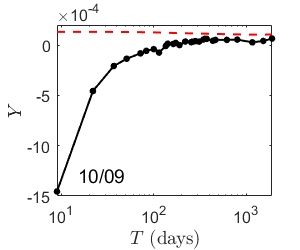}
	\caption{The implied yield curves (solid line) $Y^{(\text{med})}(t,T)$
		and (dashed line) $Y^{(\text{mkt})}(t,T)$ (PCHIP fit)
		as a function of  time to maturity $T$ for $t =$ 09 October, 2024.
		Given the range of the $y-$axis values, it is difficult to see the variation in $Y^{(\text{mkt})}(t,T)$
		with $T$ that is apparent in Fig.~\ref{fig:pchip}.
	}
	\label{fig:YMed}
\end{figure}

\begin{figure}[h!]
	\centering
	\includegraphics[width=0.7\textwidth]{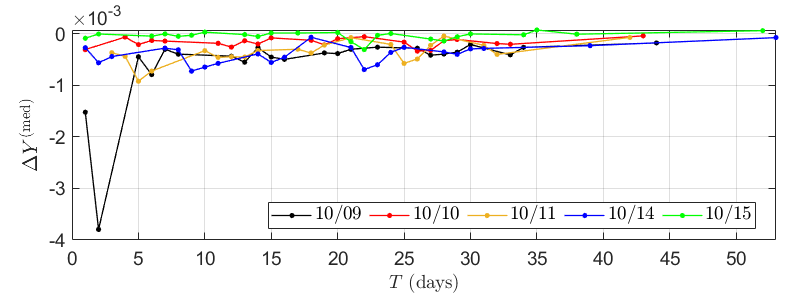}
	 \includegraphics[width=0.7\textwidth]{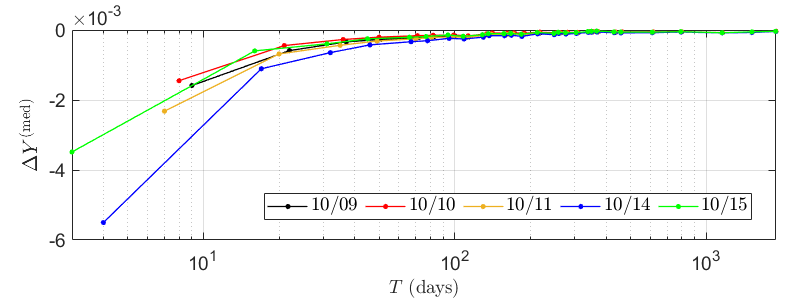}
	\caption{The dislocation $\Delta_Y^{(\text{med})}(t,T)$  between the US daily treasury par
		yield curve and 	the implied median yield curve from options on \^{}SPX for the
		(top) short--term and (bottom) long--term maturity contracts.
	}
	\label{fig:DYMED}
\end{figure}
We found it instructive to plot $Y^{(\text{med})}(t,T)$ and $Y^{(\text{mkt})}(t,T)$ separately on the same plot.
Fig.~\ref{fig:YMed} shows the plots for the short-- and long--term contracts of 09 October, 2024.
Fig.~\ref{fig:YMed_rest} in the supplementary material shows the plots for the contracts of 10--15 October, 2024.
For clearer visualization of the results, the $T-$axis is displayed using a logarithmic scale for
the long--term contracts.

For the short--term options, the median implied yield essentially hovers around a value that is negative
and less than the market yield curve for that date.
The exception is 15 October, 2024 (Fig.~\ref{fig:YMed_rest}),
where the median yield hovers around the market yield curve.
For the long--term options, the median yield monotonically rises through negative values and approaches
the long--maturity treasury yield curve rates at the longest maturity times (which approach five years).

Fig.~\ref{fig:DYMED} summarizes the results for all five days of option chains by plotting the yield dislocation
$\Delta_Y^{(\text{med})}(t,T)$ separately for the short-- and long--term contracts.
The short--term median yield curves remain below the market yield curve
while the long--term median yield curves converge consistently to the market yield curve at large maturities.

\subsection{At--the--Money Put--Call Parity Yields} \label{sec:ATM}
Examination of the variation of the implied yield at constant strike value suggested a second aggregation
method.
\begin{figure}[h!]
	\centering
	\includegraphics[width=0.8\textwidth]{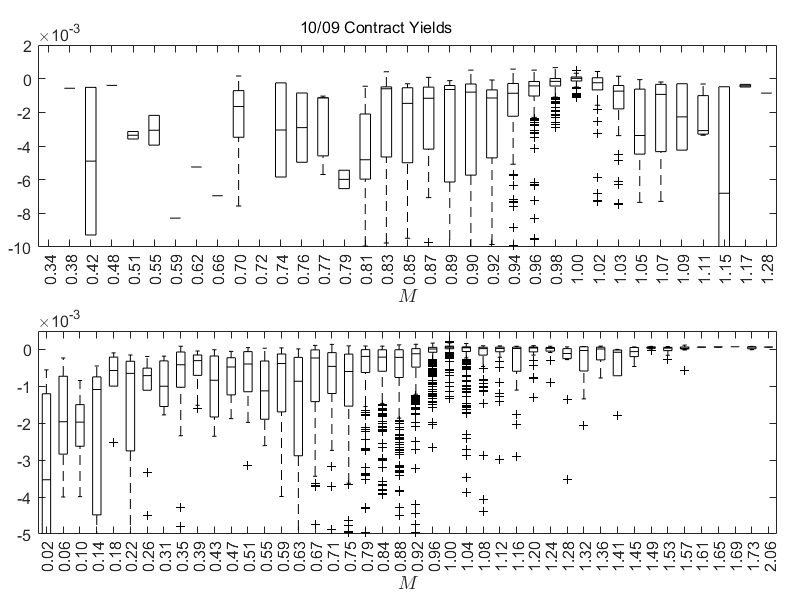}
	\caption{Box-whisker summaries of the distributions of the implied yield $Y(t,M)$ with moneyness
		$M = K / S_t$ . The panels summarize (top) short--term and (bottom) long--term contracts.
		For better representation, the negative range of the $y-$axis has been truncated
	}
	\label{fig:YMB}
\end{figure}
Fig.~\ref{fig:YMB} plots box-whisker summaries of the distributions of the implied yield $Y(t,M)$
at constant moneyness values $M = K / S_t$  for the short-- and long--term contracts for the option
chain of 09 October, 2024.\footnote{
	In order to account for the large number of strike price values, the data were grouped into strike
	price ``bins'', as would be done in a histogram.
	The strike (and hence moneyness) value characterizing each bin corresponds to the median
	of the strike values represented by the bin.
	Each box-whisker summarizes the distribution of yield values computed for the strike prices
	grouped within the bin.
}
Fig.~\ref{fig:YMB_rest} in the supplementary material  plots the box-whisker summaries for the option chains
for 10--15 October, 2024.
An examination of these distributions shows that the interquartile width narrows considerably for $M = 1$.
As $M$ moves away from unity, the distributions widen and become more negative.
Therefore, as an alternative to the median strike dislocation method~\eqref{eq:DY},
we considered the at--the--money (ATM) dislocation measure
\begin{equation}\label{eq:DY_ATM}
	\Delta_Y^{(\text{ATM})}(t,T) =  Y(t,T,M=1) - Y^{(\text{mkt})}(t,T).
\end{equation}
\begin{figure}[h!]
	\centering
	\includegraphics[width=0.3\textwidth]{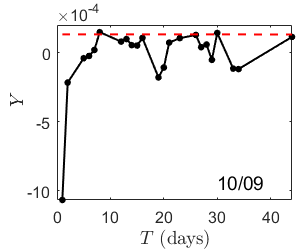}
	 \includegraphics[width=0.3\textwidth]{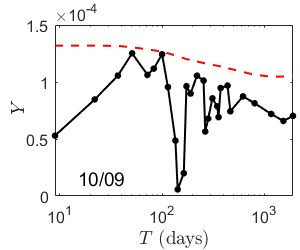}
	\caption{The yield curves (solid line) $Y^{(\text{ATM})}(t,T)$
		and (dashed line) $Y^{(\text{mkt})}(t,T)$ (PCHIP fit)
		as a function of  time to maturity $T$ for $t =$ 09 October, 2024.
	}
	\label{fig:YATM}
\end{figure}

\begin{figure}[h!]
    \centering
    \includegraphics[width=0.7\textwidth]{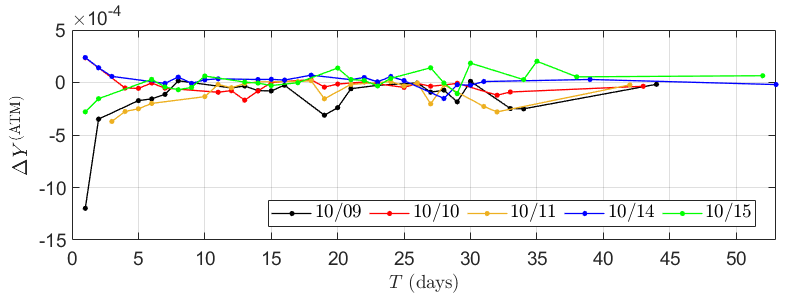}
    \includegraphics[width=0.7\textwidth]{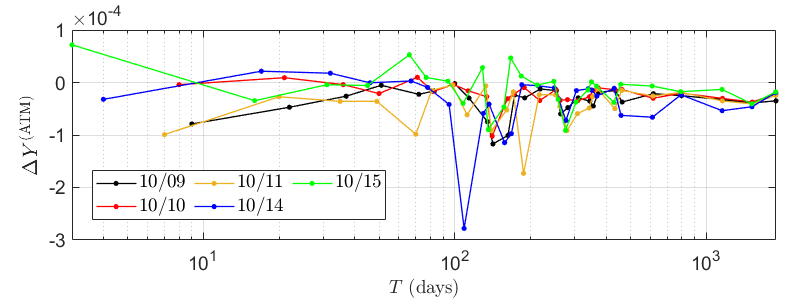}
    \caption{Dislocation between the US daily treasury par yield curve and
    		the implied at--the--money yield curve from options on \^{}SPX
    		for the (top) short--term and (bottom) long--term maturity contracts.
		Note the log scale $T$--axis in the bottom plot.
    	}
    \label{fig:DYATM}
\end{figure}

Fig.~\ref{fig:YATM} plots the yield curves $Y^{(\text{ATM})}(t,T)$ and $Y^{(\text{mkt})}(t,T)$
for the short-- and long--term contracts of 09 October, 2024.
Fig.~\ref{fig:YATM_rest} in the supplementary material shows the plots for the contracts of 10--15 October, 2024.
Fig.~\ref{fig:DYATM} displays $\Delta_Y^{(\mathrm{ATM})}(t,T)$ for each of the five trading days
$t$ during the week  9--15 October, 2024.
Unlike $Y^{(\mathrm{med})}(t,T)$, the values of $Y^{(\mathrm{ATM})}(t,T)$ tend to be closer to
the  US daily treasury par yield curve benchmark values for both short-- and long--term options.
The fluctuations tend to be slightly larger for the short--term contracts.
The curve $Y^{(\mathrm{ATM})}(t,T)$ tends to follow the par yield curve trend for large maturity times
(Figs.~\ref{fig:YATM} and \ref{fig:YATM_rest}),
although there is perhaps a slight indication that the ATM implied yield is deviating below the benchmark
yield at large maturity times (Fig.~\ref{fig:DYATM}).

Referring to Proposition~\ref{prop:BSM_TSIR}, we note that the natural option
pricing \eqref{eq:LamtT} of a unit zero bond  contains no strike price $K$.
One interpretation of a fixed, $K-$independent payout reflected by \eqref{eq:LamtT}
is that a zero--bond option is always at--the--money.
This is consistent with our findings of $Y^{(\text{ATM})}(t,T)$ providing the best fit to
the market yield curve.
\FloatBarrier

\section{Discussion}\label{sec:concl}
We have shown via use of the arbitrage--free PCP formula, that pairs of call and put options
(having common strike price and maturity date) in the \^{}SPX option chain for any date $t$ produce
a wide range of implied yield rates, which may, in fact, be negative.
The variability (Fig.~\ref{fig:YTB}) is larger for shorter term maturity dates.
For in--the--money and and out--of--the--money strike prices,  
the implied yields show significant deviation from the US daily treasury par yield curve rates,
which are commonly used as risk free rates in equity markets (Fig.~\ref{fig:YMB}).
However, implied yield rates computed from PCP using ATM option prices (Fig.~\ref{fig:DYATM})
are consistent with US daily treasury par yield curve rates.

We have argued that ATSIR models that utilize Proposition~\ref{prop:Alt_TSIR} are inconsistent in that they
implicitly assume the existence of two central banks, one dealing with equity and option prices and the other
dealing with zero bonds.
Our empirical work has demonstrate that, if the pricing of a unit zero bond is determined as if the bond were
an option (maturity $T$, payoff \$1), its price can be determined under the same risk--neutral measure
as that governing equity option markets.
Specifically, we have shown that the US daily treasury par yield rates are consistent with the
arbitrage--free yields obtained from PCP applied to at--the--money
options whose underlying, the S\&P 500 index, represents a broad component of the US equity market.
Thus, we argue, Proposition~\ref{prop:BSM_TSIR} provides the appropriate theory and
the at--the--money yield provides an appropriate TSIR framework to coherently
unite equity and fixed--income valuation.  

\normalem		

\newpage
\appendix
\renewcommand{\thefigure}{S\arabic{figure}}
\setcounter{figure}{0}
\section{Supplementary Material}\label{sec:S}
\begin{figure}[h!]
	\centering
	\includegraphics[width=0.4\textwidth]{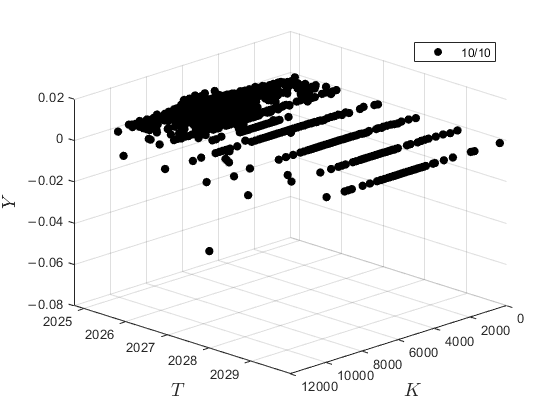}
	\includegraphics[width=0.4\textwidth]{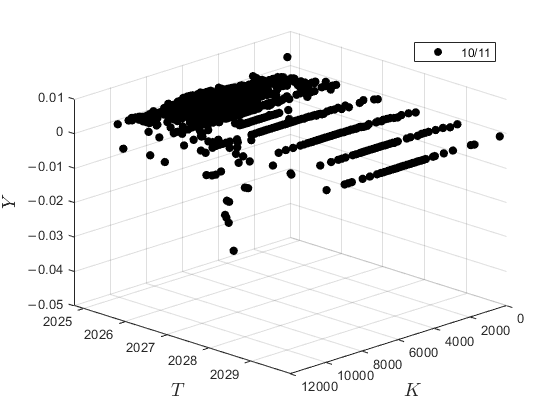}
	\includegraphics[width=0.4\textwidth]{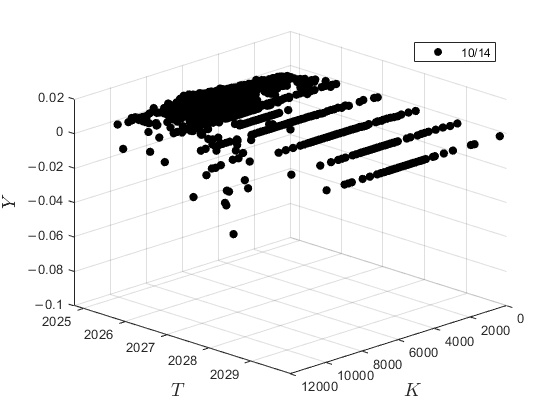}
	\includegraphics[width=0.4\textwidth]{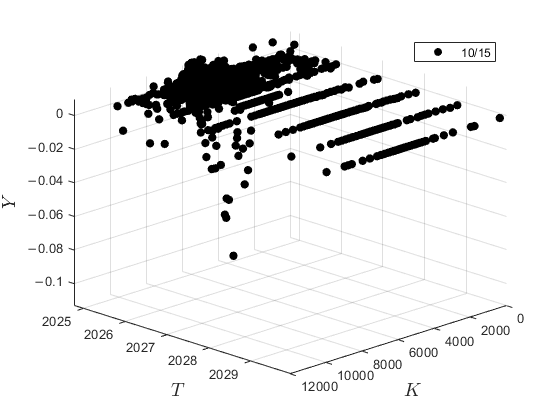}
	\caption{
		3D scatter plots of implied yields $Y(t,T,K)$
		 versus strike $K$ and maturity $T$ for the indicated trading days $t$. 
	}
	\label{fig:Y3D_rest}
\end{figure}

\begin{figure}[h!]
	\centering
	\includegraphics[width=0.49\textwidth]{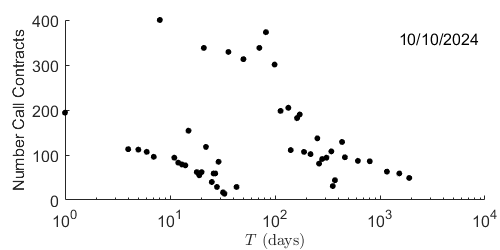}
	\includegraphics[width=0.49\textwidth]{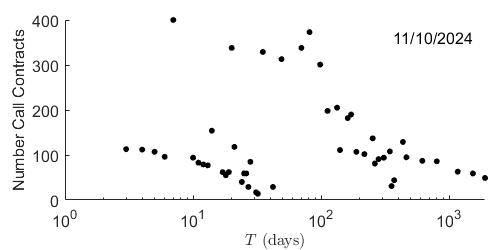}
	\includegraphics[width=0.49\textwidth]{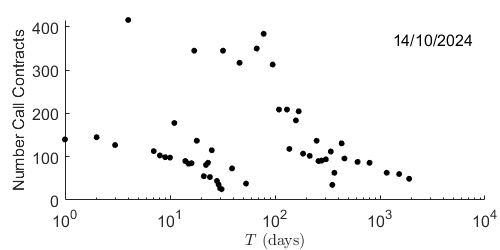}
	\includegraphics[width=0.49\textwidth]{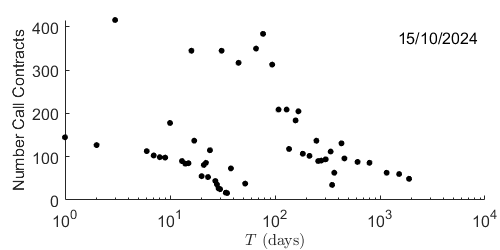}
	\caption{Two-dimensional scatter plots of the number of \^{}SPX call contracts versus time
		to maturity $T$ for the option chains of 10--15 October, 2024.}
	\label{fig:contracts_rest}
\end{figure}

\begin{figure}[h!]
	\centering
	\includegraphics[width=0.8\textwidth]{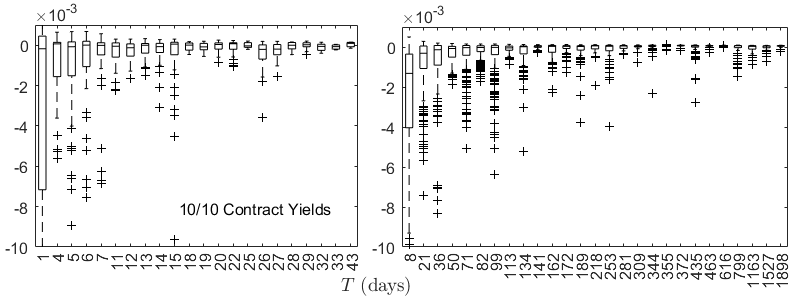}
	\caption{Box-whisker summaries of the distributions of the implied yield $Y(t,T)$ with maturity
		time $T$ (in days). The panels summarize (left) short--term and (right) long--term contracts.
		For better representation, the negative range of the $y-$axis has been truncated;
		consequently, negative outliers for small values of $T$ are not visible.
	}
	\label{fig:YTB_rest}
\end{figure}
\begin{figure}[h!]\ContinuedFloat
	\centering
	\includegraphics[width=0.8\textwidth]{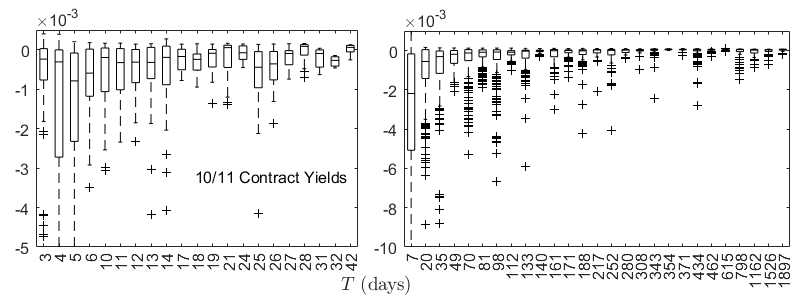}
	\includegraphics[width=0.8\textwidth]{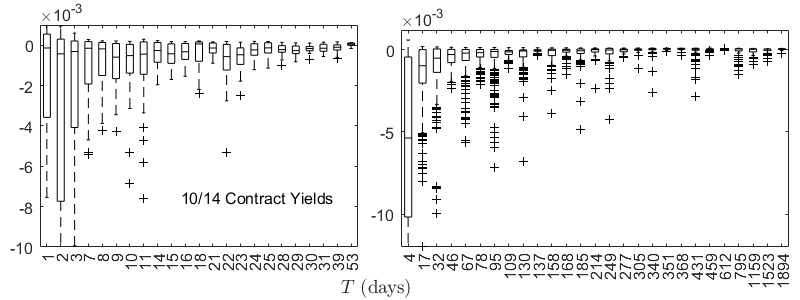}
	\includegraphics[width=0.8\textwidth]{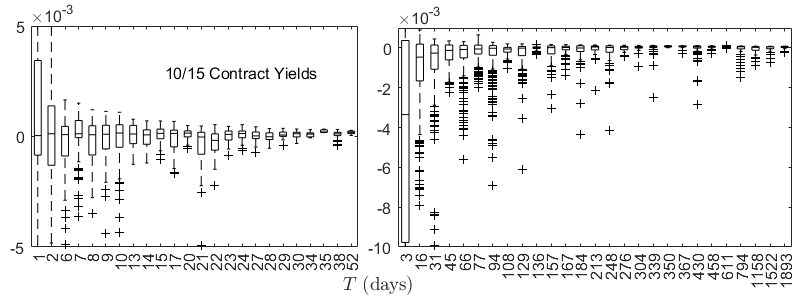}
	\caption{
		(cont.)
	}
\end{figure}

\begin{figure}[h!]
	\centering
	\includegraphics[width=0.24\textwidth]{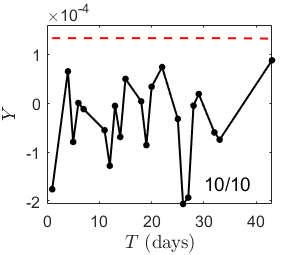}
	\includegraphics[width=0.24\textwidth]{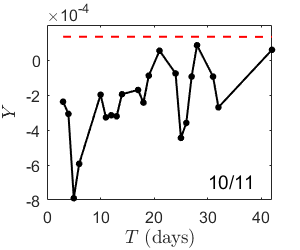}
	\includegraphics[width=0.24\textwidth]{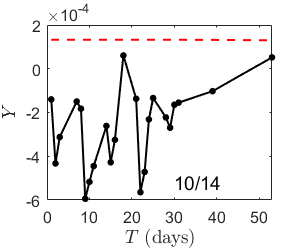}
	\includegraphics[width=0.24\textwidth]{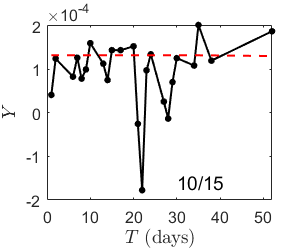}
	 \includegraphics[width=0.24\textwidth]{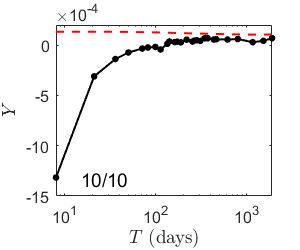}
	 \includegraphics[width=0.24\textwidth]{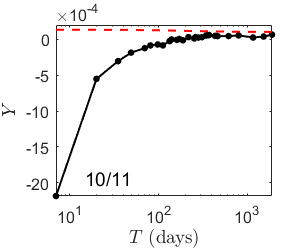}
	 \includegraphics[width=0.24\textwidth]{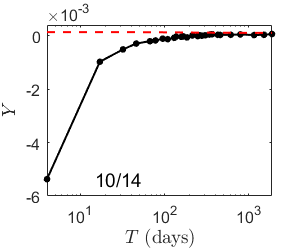}
	 \includegraphics[width=0.24\textwidth]{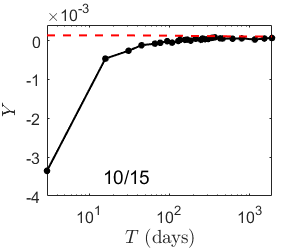}
	\caption{(solid line) $Y^{(\text{med})}(t,T)$ and (dashed line) $Y^{(\text{mkt})}(t,T)$ (PCHIP fit)
		as a function of time to maturity $T$ for $t =$ 10--15 October, 2024.
		(top row) Short--term contracts. (bottom row) Long--term contracts.
		Given the range of the $y-$axis values, it is difficult to see the variation in $Y^{(\text{mkt})}(t,T)$
		with $T$ for large $T$ that is apparent in Fig.~\ref{fig:pchip}.
	}
	\label{fig:YMed_rest}
\end{figure}

\begin{figure}[h!]
	\centering
	\includegraphics[width=0.75\textwidth]{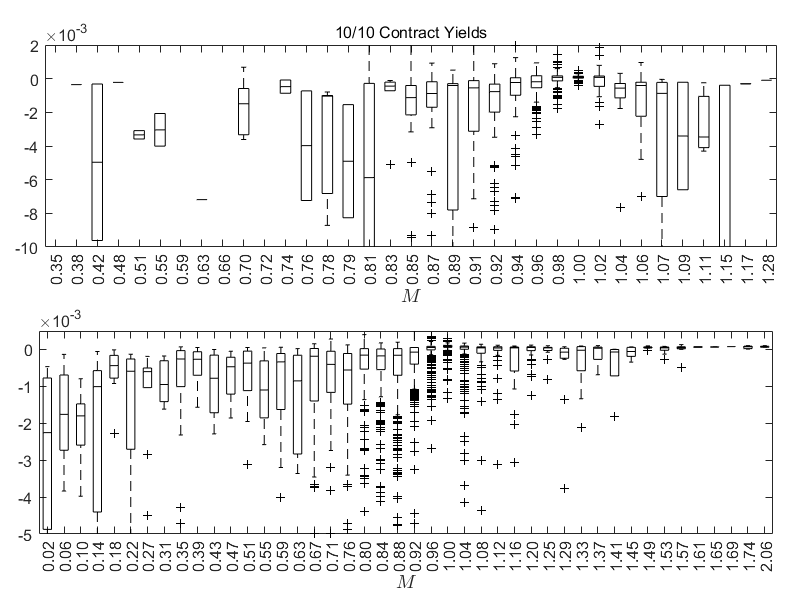}
	\caption{Box-whisker summaries of the distributions of the implied yield $Y(t,M)$ with moneyness
		$M = K / S_t$  for 10 October, 2024. The panels summarize (top) short--term and (bottom) long--term contracts.
		For better representation, the negative range of the $y-$axis has been truncated
	}
	\label{fig:YMB_rest}
\end{figure}
\begin{figure}[h!]\ContinuedFloat
	\centering
	\includegraphics[width=0.75\textwidth]{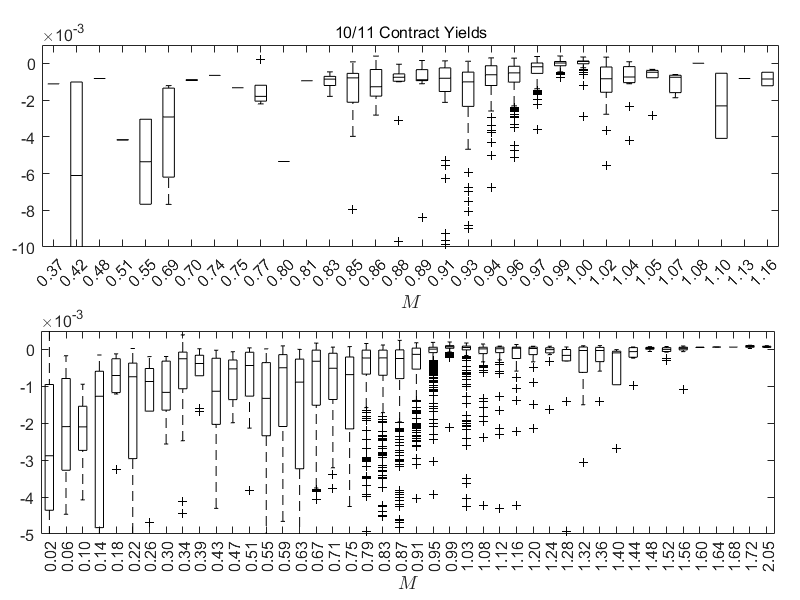}
	\includegraphics[width=0.75\textwidth]{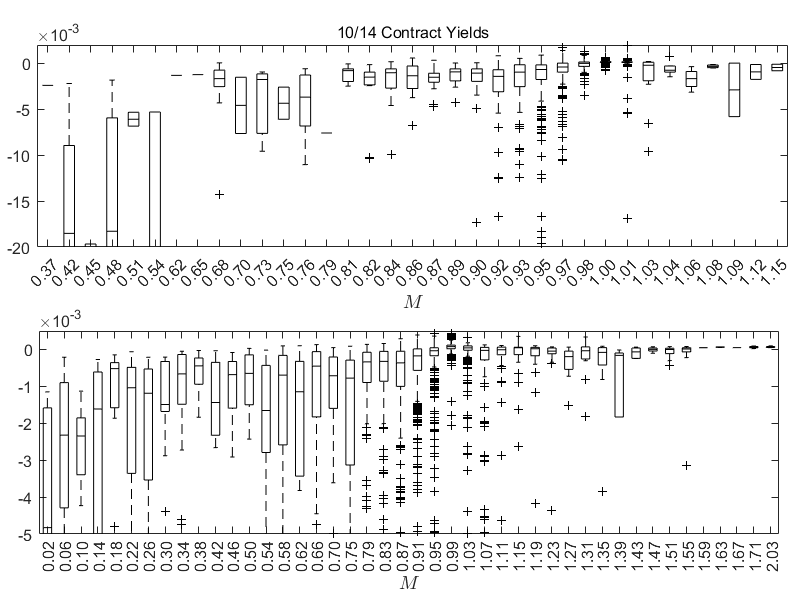}
	\caption{(cont.)}
\end{figure}
\begin{figure}[h!]\ContinuedFloat
	\centering
	\includegraphics[width=0.75\textwidth]{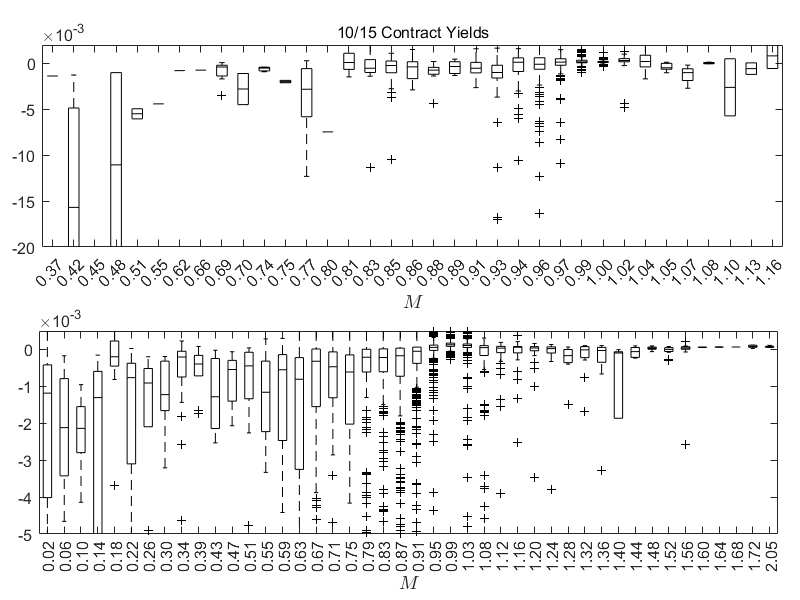}
	\caption{(cont.)}
\end{figure}

\begin{figure}[h!]
	\centering
	\includegraphics[width=0.24\textwidth]{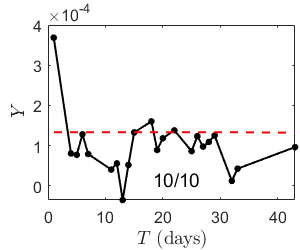}
	\includegraphics[width=0.24\textwidth]{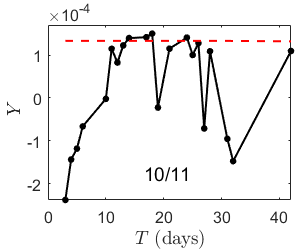}
	\includegraphics[width=0.24\textwidth]{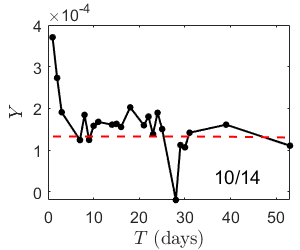}
	\includegraphics[width=0.24\textwidth]{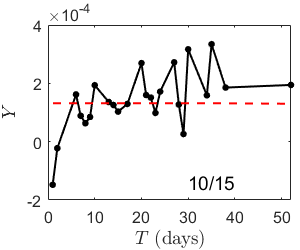}
	 \includegraphics[width=0.24\textwidth]{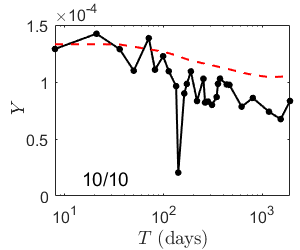}
	 \includegraphics[width=0.24\textwidth]{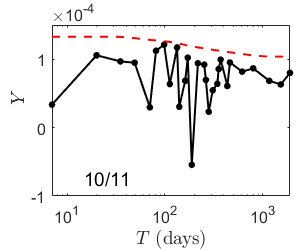}
	 \includegraphics[width=0.24\textwidth]{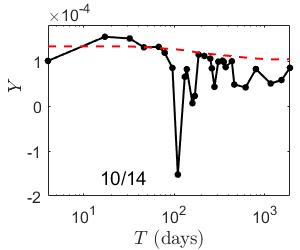}
	 \includegraphics[width=0.24\textwidth]{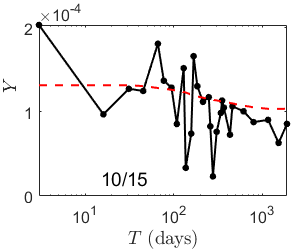}
	\caption{(solid line) $Y^{(\text{ATM})}(t,T)$ and (dashed line) $Y^{(\text{mkt})}(t,T)$ (PCHIP fit)
		as a function of time to maturity $T$ for $t =$ 10--15 October, 2024.
		(top row) Short--term contracts. (bottom row) Long--term contracts.
	}
	\label{fig:YATM_rest}
\end{figure}

\end{document}